\documentclass{ws-ijmpa}
\def\halftext{.450\textwidth}

\newcommand{\BR}{\mathcal{B}}

\newcommand{\gev}{\,\mbox{GeV}}

\newcommand{\ra}{\rightarrow}

\newcommand{\psp}{\psi(2S)}

\newcommand{\pspto}{\psi^{\prime}\to}
\newcommand{\jpsi}{J/\psi}
\newcommand{\jpsito}{J/\psi\to}
\newcommand{\pspp}{\psi(3770)}

\newcommand{\EE}{e^+e^-}
\newcommand{\EETO}{e^+e^-\to}
\newcommand{\pip}{\pi^+}
\newcommand{\pim}{\pi^-}
\newcommand{\piz}{\pi^0}

\newcommand{\rhoto}{\rho(2150)}

\newcommand{\rhopi}{\rho\pi}

\newcommand{\rhotopi}{\rhoto\pi}

\newcommand{\bfg}{\begin{figure}}
\newcommand{\efg}{\end{figure}}
\newcommand{\bitm}{\begin{itemize}}
\newcommand{\eitm}{\end{itemize}}
\newcommand{\bnum}{\begin{enumerate}}
\newcommand{\enum}{\end{enumerate}}
\newcommand{\btbl}{\begin{table}}
\newcommand{\etbl}{\end{table}}
\newcommand{\btbu}{\begin{tabular}}
\newcommand{\etbu}{\end{tabular}}
\newcommand{\beqns}{\begin{eqnarray*}}
\newcommand{\eeqns}{\end{eqnarray*}}
\begin{document}

\markboth{Zheng Wang} {Recent results of $\psi(2S)$ decays from
BES}

%
\catchline{}{}{}{}{}
%

\title{Recent Results of $\psi(2S)$ Decays from BES}

\author{\footnotesize Zheng Wang\\
(For the BES Collaboration)
}

\address{China Center for Advanced Science and Technology(CCAST),\\
Beijing 100080, People's Republic of China\\
E-mail:~wangz@mail.ihep.ac.cn\\
The International Conference on QCD and Hadronic Physics\\
16-20~Jun., 2005, Peking University, Beijing, China
 }

%

\maketitle


\begin{abstract}
Results of many $\psi(2S)$ hadronic decays and the results of
$\psi(2S)$ radiative/hadronic transitions from BES collaboration
in the past year are presented. Measurement of $\BR(\jpsi\ra K^0_S
K^0_L)$ and the preliminary result of searching for the decay
$\psi(3770)\ra\rho\pi$ are also reported.

\end{abstract}

\section{Introduction}
The Beijing Spectrometer (BES) is a general purpose solenoidal
detector running at the Beijing Electron-Positron Collider (BEPC)
storage ring. BEPC operates in the center of mass energy range
from 2 to 5 GeV with a luminosity at the $\psp$ energy of
approximately $1\times 10^{31}~\hbox{cm}^{-2}\hbox{s}^{-1}$. The
numbers/luminosities of $\psi(2S)$, $\jpsi$, $\pspp$ and the
continuum events ($\sqrt{s}=3.65\gev$) taken with the BES
detector, are listed in Table~\ref{data}. BES (BESI) is described
in detail in Ref.~\refcite{bes}, and the upgraded BES detector
(BESII) is described in Ref.~\refcite{bes2}.
\begin{table}[h]
\tbl{\label{data}BES statistics in use for analyses}
{\begin{tabular}{@{}c|c|c|c|c|c@{}}\hline
Detector config.   & \multicolumn{4}{|c|}{BESII}  & BESI\\
\hline
Ecm (GeV)& 3.686 & 3.097 & 3.773 & 3.650  & 3.686\\
\hline Tot. Num.(millon)  &$14.0\pm 0.6$ &$57.7\pm 2.7$ &
                   &
                  & $3.79\pm 0.31$ \\
or Int.$\mathcal{L}$($pb^{-1}$) & & &$17.3\pm 0.5$ &$6.42\pm 0.24$ & \\
\hline
\end{tabular}}
\etbl
\section{Results of $\psi(2S)$ hadronic final state decays}
From perturbative QCD (pQCD), it is expected that both $\jpsi$ and
$\psp$ decaying into light hadrons are dominated by the
annihilation of $c\bar{c}$ into three gluons or one virtual
photon, with a width proportional to the square of the wave
function at the origin~cite. This yields the pQCD ``12\% rule'',
\begin{equation}
 Q_h =\frac{{\cal B}_{\pspto h}}{{\cal
B}_{\jpsito h}} =\frac{{\cal B}_{\pspto \EE}}{{\cal B}_{\jpsito
\EE}} \approx 12\%.
\end{equation}
In order to test the above rule, systematic studies of $\psp$
exclusive hadronic decays have been carried out at the BES.
Table~\ref{had-result} summarizes the results of branching
fraction measurements for all decays modes of the $\psp$ studied.
The table includes the data for the corresponding $\jpsi$ decays
as well as the ratios of branching fractions of the $\psp$ to the
$\jpsi$. Some interesting features are found: At BES, a Partial
Wave Analysis (PWA) is carried out for the $\pip\pim\piz$ final
state using the helicity amplitude method. $\psp\ra\rho(770)\pi$
is observed. In the $\pi\pi$ mass spectrum, a high mass
enhancement with mass round $2.15\gev$ is observed, seen in
Figure~\ref{fit}. Attributing this enhancement to the $\rho(2150)$
resonance, the branching fraction is measured to be $\BR(\pspto
\rho(2150)\pi \to \pi^+ \pi^- \pi^0) = (19.4 \pm 2.5
^{+11.5}_{-3.4}) \times 10^{-5}$; The ratio $\frac{\BR(\psp\ra
K^{*+}K^-+c.c.)}{\BR(\psp\ra
K^{*0}\bar{K^0}+c.c.)}=4.6^{+2.9}_{-2.2}$ from BES shows a large
isospin violation between the charged and neutral modes of
$\psp\ra K^*(892)\bar{K}+c.c.$ decays; Compared with $\jpsi\ra
K^0_S K^0_L$~\cite{jpsikskl}, BES measurement of $\BR(\psp\ra
K^0_S K^0_L)$ is enhanced relative to the ``12\% rule''. More
details on these analyses may be found in
Refs.~\refcite{jpsikskl}, and~\refcite{hadron}.

\begin{table}[h]
\tbl{\label{had-result}Branching fractions measured for the $\psp$
decays. Results for the corresponding $\jpsi$ branching fractions
from PDG  are also given as well as the ratios
$Q_h=\BR(\psp)/\BR(\jpsi)$. All limits are at 90\% confidence
level.  } {\begin{tabular}{@{}l|l|l|l|l@{}} \hline
 Mode &
Channel & $\BR(\psp)~(\times10^{-5})$
     & $\BR(\jpsi)~(\times10^{-4})$~\cite{pdg}
     & $Q_h~(\%)$ \\
\hline
       &  $\rho\pi$   & $5.1\pm0.7\pm1.1$ & $127\pm9$
                      &$0.40\pm0.11$\\
\cline{2-5}
     & $K^{\star +}K^-+c.c.$  & $2.9^{+1.3}_{-1.7}\pm0.4$ &
       $ 50\pm 4 $    & $0.59^{+0.27}_{-0.36}$ \\
 \cline{2-5}
       & $K^{\star 0}\bar{K^0}+c.c.$  & $13.3^{+2.4}_{-2.8}\pm1.7$ &
       $42\pm 4$ & $3.2\pm0.8$\\
 \cline{2-5}
       & $\phi\piz$  & $<0.41$ & $<0.068$ & --\\
  \cline{2-5}
       & $\phi\eta$  & $3.3\pm1.1\pm0.5$ & $6.5\pm0.7$
                     & $5.1\pm1.9$\\
  \cline{2-5}
   VP    & $\phi\eta^{\prime}$  & $2.8\pm1.5\pm0.6$  & $3.3\pm0.4$
                     & $8.5\pm5.0$\\
  \cline{2-5}
       & $\omega\eta$  & $<3.2$ & $15.8\pm1.6$
                     & $<2.0$\\
   \cline{2-5}
       & $\omega\eta^{\prime}$  & $3.1^{+2.4}_{-2.0}\pm0.7$ &
       $1.67\pm0.25$ & $19^{+15}_{-13}$\\
    \cline{2-5}
       & $\omega\piz$  &$1.87^{+0.68}_{-0.62}\pm0.28$ &
       $4.2\pm0.6$ & $4.4^{+1.8}_{-1.6}$ \\
       \cline{2-5}
       & $\rho\eta$  & $1.78^{+0.67}_{-0.62}\pm0.17$ &
       $1.93\pm0.23$ &  $9.2^{+3.6}_{-3.3}$\\
        \cline{2-5}
       & $\rho\eta^{\prime}$  & $1.87^{+1.64}_{-1.11}\pm0.33$ &
       $1.05\pm0.18$ & $17.8^{+15.9}_{-11.1}$\\
                     \hline
      & $K^0_SK^0_L$  & $5.24\pm0.47\pm0.48$ &
      $18.2\pm1.4$~\cite{jpsikskl}
                     &$28.8\pm3.7$\\
      \cline{2-5}
 PP and   & $\pip\pim\piz$  &$18.1\pm1.8\pm1.9$
          & $210\pm12$~\cite{j3pi}
                     &$0.86\pm0.13 $\\
 \cline{2-5}
 multi-body  & $p\bar{p}\piz$  & $13.2\pm1.0\pm1.5$ &
             $10.9\pm0.09$ & $12.1\pm1.9$\\
  \cline{2-5}
 decays  & $p\bar{p}\eta$  & $5.8\pm1.1\pm0.7$ & $2.09\pm0.18$
                     & $2.8\pm0.7$\\
  \cline{2-5}
       & $3(\pip\pim)$  & $54.5\pm4.2\pm8.7$ & $40\pm20$
                     & $14\pm8$\\
  \hline
  \end{tabular}}
  \end{table}

BES also searchs for the Non-$D\bar{D}$ decay of $\psi(3770)\ra
\rhopi$ using a data sample of $(17.3\pm 0.5)~pb^{-1}$ taken at
the center-of-mass energy of 3.773 GeV~\cite{ppcon}. No $\rhopi$
signal is observed, and the upper limit of the cross section is
measured to be $\sigma(\EETO \rhopi)<6.0~pb$ at 90\% C. L.
Considering the interference between the continuum amplitude and
the $\psi(3770)$ resonance amplitude, the branching fraction of
$\psi(3770)$ decays to $\rho\pi$ is determined to be
$\BR(\psi(3770)\ra\rho\pi)\in(6.0\times10^{-6},~2.4\times10^{-3})$
at 90\% C. L., as shown in Figure~\ref{sborn}. This is in
agreement with the prediction of the $S$- and $D$-wave mixing
scheme of the charmonium states for solving the ``$\rhopi$
puzzle'' between $\jpsi$ and $\psi(2S)$
decays~\cite{rosnersd,wympspp}.

\begin{figure}[htb]
\parbox{\halftext}{\psfig{file=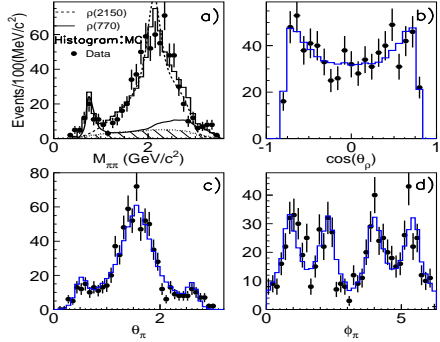,width=\halftext,height=4.5cm}
\caption{Comparison between data (dots with error bars) and the
final fit (solid histograms) in $\psp\ra\pip\pim\piz$ PWA. (a) two
pion invariant mass, with a solid line for the $\rho(770)\pi$, a
dashed line for the $\rhotopi$, and a hatched histogram for
background; (b) the $\rho$ polar angle in the $\psp$ rest frame;
and (c) and (d) for the polar and azimuthal angles of the
designated $\pi$ in $\rho$ helicity frame.}\label{fit}}
         \hspace{2mm}
\parbox{\halftext}{\psfig{file=ang_cro5.epsi,width=\halftext,height=5.0cm}
\caption{Restriction on $\BR(\pspp\ra\rho\pi)$ and $\phi$, the
relative phase between $\pspp$ strong and electromagnetic decay
amplitudes, from the measurement of $\rho\pi$ at the $\pspp$ peak
in this experiment. The hatched area indicate the physical region
at 90\% C. L.}\label{sborn}}
\end{figure}
\section{Results of $\psi(2S)$ radiative and hadronic transitions}
We report on the analysis of $\psp\ra\piz\jpsi$, $\eta\jpsi$, and
$\gamma\chi_{c1,2}$ decays based on a sample of $14.0\times 10^6$
$\psp$ events collected with the BESII detector. Here $\jpsi$ is
reconstructed by $\EE$ of $\mu^+\mu^-$. Another analysis, based on
a sample of approximately $4\times 10^6~\psp$ events obtained with
the BESI detector, is used for measuring branching fractions for
the inclusive decay $\psp\ra$ anything $\jpsi$, and the exclusive
processes for the cases where $X~=~\eta$ and $X~=~\pi\pi$, as well
as the cascade precesses
$\psp\ra\gamma\chi_{c0/1/2}\ra\gamma\gamma\jpsi$. The branching
fractions and the ratios of the branching fractions are shown in
Tables~\ref{emtab1} and \ref{emtab2}, along with the PDG
value~\cite{pdg}. Details on these analyses may be found in
Refs.~\refcite{em1} and \refcite{em2}.

\begin{table}[h]
\tbl{Results from $\psp\ra\gamma\gamma\jpsi$ (BESII data).}
{\begin{tabular}{@{}c|cc|cc@{}} \hline Channel &
\multicolumn{2}{|c|}{$\piz\jpsi$} &
\multicolumn{2}{|c}{$\eta\jpsi$}  \\
\hline
$\BR$ (\%) & \multicolumn{2}{|c|}{$0.143\pm0.014\pm0.013$}
& \multicolumn{2}{|c}{$2.98\pm0.09\pm0.23$}\\
PDG (\%)~\cite{pdg} & \multicolumn{2}{|c|}{$0.096\pm0.021$} &
  \multicolumn{2}{|c}{$3.16\pm0.22$}\\
\hline Channel & \multicolumn{2}{|c}{$\gamma\chi_{c1}$}
&\multicolumn{2}{c}{$\gamma\chi_{c2}$}\\
 \hline $\BR$ (\%) & \multicolumn{2}{|c|}{$8.90\pm0.16\pm1.05$} &
  \multicolumn{2}{|c}{$8.02\pm0.17\pm0.94$}\\
PDG (\%)~\cite{pdg} & \multicolumn{2}{|c|}{$8.4\pm0.6$} &
  \multicolumn{2}{|c}{$6.4\pm0.6$}\\
 \hline
\end{tabular}}
\label{emtab1}
\end{table}
\begin{table}[h]
\tbl{Branching ratios and branching fractions (BESI data).
PDG04-exp results are single measurements or averages of
measurements, while PDG04-fit are results of their global fit to
many experimental measurements. The BES results in the second half
of the table are calculated using the PDG04 value of
$\BR_{\pi\pi}=\BR(\psp\ra\jpsi\pip\pim)=(31.7\pm1.1)\%$.}
{\begin{tabular}{@{}l|c|c|c@{}} \toprule Case & This exp. &
PDG04-exp & PDG04-fit \\
\hline
 $\BR(\jpsi(\hbox{anything}))/\BR_{\pi\pi}$ &
$1.867\pm0.026\pm0.055$ & $2.016\pm0.150$
& $1.821\pm0.036$  \\
$\BR(\jpsi\piz\piz)/\BR_{\pi\pi}$ & $0.570\pm0.009\pm0.026$ & -
& $0.59\pm0.05$  \\
$\BR(\jpsi\eta)/\BR_{\pi\pi}$ & $0.098\pm0.005\pm0.010$ &
$0.091\pm0.021$
& $0.100\pm0.008$  \\
$\BR(\gamma\chi_{c1})\BR(\chi_{c1}\ra\gamma\jpsi)/\BR_{\pi\pi}$ &
$0.126\pm0.003\pm0.038$ & $0.085\pm0.021$
& $0.084\pm0.006$  \\
$\BR(\gamma\chi_{c2})\BR(\chi_{c2}\ra\gamma\jpsi)/\BR_{\pi\pi}$&
$0.060\pm0.000\pm0.028$ & $0.039\pm0.012$
& $0.041\pm0.003$  \\
\hline $\BR(\jpsi(\hbox{anything}))~(\%)$ & $59.2\pm0.8\pm2.7$ &
$55\pm7$
& $57.6\pm2.0$  \\
$\BR(\jpsi\piz\piz)~(\%)$ & $18.1\pm0.3\pm1.0$ & -
& $18.8\pm1.2$  \\
$\BR(\jpsi\eta)~(\%)$ & $3.11\pm0.17\pm0.31$ & $2.9\pm0.5$
& $3.16\pm0.22$  \\
$\BR(\gamma\chi_{c1})\BR(\chi_{c1}\ra\gamma\jpsi)~(\%)$ &
$4.0\pm0.1\pm1.2$ & $2.66\pm0.16$
& $2.67\pm0.15$  \\
$\BR(\gamma\chi_{c2})\BR(\chi_{c2}\ra\gamma\jpsi)~(\%)$&
$1.91\pm0.01\pm0.86$ & $1.20\pm0.13$ & $1.30\pm0.08$
\\
\hline
\end{tabular}}
\label{emtab2}
\end{table}
\section*{Acknowledgments}
I wish to acknowledge the efforts of my BES colleagues on all the
results presented here. I also want to thank the organizers for
the opportunity to present these results at the International
conference on QCD and hadronic Physics at Peking University.

\end{document}